\documentclass[12pt,english,floatfix,superscriptaddress,aps,prd,preprint]{revtex4}
\usepackage[utf8]{inputenc}
\usepackage{amsmath}
\usepackage{amssymb}
\usepackage{amsbsy}
\usepackage{amsfonts}
\usepackage{amsopn}
\usepackage{amstext}
\usepackage{graphicx}
\usepackage{amssymb}
\usepackage{amsfonts}
\usepackage{amsmath}
\usepackage{amsmath,amsthm,amsfonts,amssymb}
\usepackage[mathcal]{eucal}
\usepackage{mathrsfs}
\usepackage{graphicx}
\usepackage{color}
\usepackage{esint}
\usepackage[dvips]{epsfig}
\usepackage[dvips]{graphicx}
\usepackage{float}
\usepackage{units}
\usepackage{textcomp}

\usepackage{hyperref}
\usepackage{slashed}

\newcommand{\ie}{\begin{equation}}
\newcommand{\fe}{\end{equation}}
\newcommand{\se}{\begin{eqnarray}}
\newcommand{\ff}{\end{eqnarray}}

\begin{document}

\title{Bound state solutions of the Dirac oscillator in an Aharonov-Bohm-Coulomb system}


\author{R. R. S. Oliveira}
\email{rubensrso@fisica.ufc.br}
\affiliation{Universidade Federal do Cear\'a (UFC), Departamento de F\'isica,\\ Campus do Pici, Fortaleza - CE, C.P. 6030, 60455-760 - Brazil.}

\author{R. V. Maluf}
\email{r.v.maluf@fisica.ufc.br}
\affiliation{Universidade Federal do Cear\'a (UFC), Departamento de F\'isica,\\ Campus do Pici, Fortaleza - CE, C.P. 6030, 60455-760 - Brazil.}


\author{C. A. S. Almeida}
\email{carlos@fisica.ufc.br}
\affiliation{Universidade Federal do Cear\'a (UFC), Departamento de F\'isica,\\ Campus do Pici, Fortaleza - CE, C.P. 6030, 60455-760 - Brazil.}

\date{\today}

\begin{abstract}
In this work, we study of the (2+1)-dimensional Dirac oscillator in the presence of a homogeneous magnetic field in an Aharonov-Bohm-Coulomb system. To solve our system, we apply the $left$-$handed$ and $right$-$handed$ projection operators in the Dirac oscillator to obtain a biconfluent Heun equation. Next, we explicitly determine the energy spectrum for the bound states of the system and their exact dependence on the cyclotron frequency $\omega_c$ and on the parameters $Z$ and $\Phi_{AB}$ that characterize the Aharonov-Bohm-Coulomb system. As a result, we observe that by adjusting the frequency of the Dirac oscillator to resonate with the cyclotron half-frequency the energy spectrum reduces to the rest energy of the particle. Also, we determine the exact eigenfunctions, angular frequencies, and energy levels of the Dirac oscillator for the ground state ($n=1$) and the first excited state ($n=2$). In this case, the energy levels do not depend on the homogeneous magnetic field, and the angular frequencies are real and positive quantities, increase quadratically with the energy and linearly with $\omega_{c}$.
\end{abstract}

\maketitle

\section{Introduction}

Since the birth of quantum mechanics, the existence of relevant physical systems that have exact analytical solutions is limited to a few cases of interest. Analytical solutions are essential for a better understanding of physical phenomena and allow accurate comparison between theory and experiment. In the context of relativistic quantum mechanics, an analytically soluble model widely explored in the literature is the so-called Dirac oscillator (DO), first introduced by Ito et al. \cite{Ito}. This model was named after Moshinsky and Szczepaniak \cite{Moshinsky}, who introduced a relativistic version of the harmonic oscillator into the Dirac equation (DE). In this system, the coupling term is linear with respect to the position vector, satisfying the Lorentz invariance and, in non-relativistic limit, it reduces to the Hamiltonian that is commonly used for the quantum harmonic oscillator added with a strong spin-orbit coupling \cite{Martinez}.

The Dirac oscillator equation is obtained through the non-minimal substitution ${\bf p}\rightarrow {\bf p}-i m\omega \beta {\bf r}$, where $m$ is the rest mass of the particle, $\omega $ is the angular frequency of the oscillator, {\bf{r}} is the position vector, and $\beta$ is the usual Dirac matrix \cite{Moshinsky}. From a physical viewpoint, this non-minimal substitution can be interpreted as an interaction of the anomalous magnetic moment of a Dirac neutral fermion with an electric field generated by a uniformly charged dielectric sphere \cite{Martinez}. Since it was proposed in the literature, several works on DO were carried out in different contexts. For example, in the studies of thermodynamic properties \citep{Pacheco,Boumali2013,Hassanabadi2015,Sargolzaeipor}, mathematical physics \cite{Hatami, Maluf,Andrade,Lange,Szmytkowski,Quesne2006}, quantum chromodynamics \cite{Moshinsky2,DeLange}, quantum optics \cite{Zitterbewegung,Longhi}, graphene physics \cite{Quimbay2013cond-mat.,Boumali,Belouad,Neto2016}, and so on.  Recently, the first experimental realization of the one-dimensional Dirac oscillator was obtained by Franco-Villafa\~{n}e et al. \cite{Franco}. The experimental setup is based on a microwave system consisting of a chain of coupled dielectric disks, such that the coupling between the disks can be adjusted to obtain a spectrum equivalent to the DO.

Furthermore, in recent years, several works have investigated the dynamics of relativistic and non-relativistic Dirac fermions interacting with the vector potential of the Aharonov-Bohm (AB) effect and the Coulomb potential. A system described by the combination of these two potentials is so-called Aharonov-Bohm-Coulomb (ABC) system \cite{Hagen,Yoo}. For instance, the ABC system is studied in connection with the Feynmann path integrals \cite{Lin1999,Bornales}, scattering process \cite{Lin2000}, nongauge potentials \cite{Hagen1993}, self-adjoint extension approach \cite{Park}, magnetic monopole \cite{Villalba1995,Hai}, spontaneous creation of fermions pairs and vacuum polarization \cite{Khalilov2009}, graphene quantum ring \cite{Jung}, supercritical Coulomb impurity \cite{Nishida}, and so on.

In this work, we investigate the relativistic quantum dynamics of the (2+1)-dimensional DO in the presence of a homogeneous magnetic field in an ABC system. We apply the biconfluent Heun equation (BHE) formalism \cite{Ronveaux} to obtain the relativistic energy levels for the DO, as well as the eigenfunctions and the angular frequencies. This formalism is an efficient method for getting bound state solutions, energy spectrum and angular frequencies for (2+1)-dimensional relativistic quantum systems subject to the harmonic oscillator potentials  \cite{Vitoria,Medeiros,Bakke}. In our case, we explicitly determine the dependence of the energy spectrum on the cyclotron frequency $\omega_c$ and the parameters $Z$ and $\Phi_{AB}$ that characterize the ABC system. Moreover, we evaluated the exact eigenfunctions, angular frequencies, and energy levels of the DO for the ground state ($n=1$) and the first excited state ($n=2$). As a result, we show that the energy levels do not depend on the homogeneous magnetic field, and the angular frequencies are real and positive quantities, increase quadratically with the energy and linearly with $\omega_{c}$.

The work is organized as follows. In Section \ref{sec2}, we introduce the DO in polar coordinates in the presence of a homogeneous magnetic field in an ABC system. Next, we apply the projection operators $left$-$handed$ and $right$-$handed$ in the DO and we obtain a biconfluent Heun equation. In Section \ref{sec3}, we explicitly determined the energy spectrum, the eigenfunctions and the angular frequencies for the bound states of the system. As an example, we obtain precisely the eigenfunctions and angular frequencies of the DO for the ground state and first excited state. In Section \ref{conclusion}, we present our final comments and conclusions.

\section{Dirac Oscillator in the presence of an homogeneous magnetic field in an Ahharonov-Bohm-Coulomb system \label{sec2}}

The covariant (2+1)-dimensional DE in the presence of an external electromagnetic field $A_\mu$ reads as follows (in Gaussian system and natural units $\hbar=c=1$) \cite{Greiner}
\ie (\gamma^{\mu}\Pi_{\mu}-m_{0})\Psi(t,{\bf r})=0, \ \ (\mu=0,1,2),
\label{mass1}\fe
where $\gamma^{\mu}$ are the Dirac gamma matrices, $\Pi_{\mu}=p_\mu-qA_\mu$ is the kinetic momentum operator and $q$, $p_{\mu}$ and $\Psi(t,{\bf r})$ are the electric charge of the fermion, canonical momentum operator and two-component Dirac spinor, respectively.  
Now, we define the following projection operators $left$-$handed$ and $right$-$handed$ \cite{Auvil,Kaplan}
\ie P_{L}=\frac{1}{2}(\mathbb{I}-\gamma^{5}), \ \ P_{R}=\frac{1}{2}(\mathbb{I}+\gamma^{5}),
\label{operators1}\fe
which satisfy the properties $P^{2}_{L}=P_{L}, P^{2}_{R}=P_{R},\{P_{L},P_{R}\}=0, P_{L}+P_{R}=\mathbb{I},P_{R}\gamma^{\mu}=\gamma^{\mu}P_{L}$, where $\gamma^{5}=\gamma_{5}=\sigma_{1}$ is called the chirality operator. Applying $P_L$ in the Eq. \eqref{mass1} and defining the $left$-$handed$ and $right$-$handed$ spinors as $\Psi_L(t,{\bf r})=P_L\Psi(t,{\bf r})$ and $\Psi_R(t,{\bf r})=P_R\Psi(t,{\bf r})$, we get
\ie \Psi_L(t,{\bf r})=(m_{0})^{-1}\gamma^{\mu}\Pi_{\mu}\Psi_R(t,{\bf r}).
\label{mass2}\fe

The above relation allows us to write the original spinor in the form 
\ie \Psi(t,{\bf r})=(m_{0})^{-1}[\gamma^{\mu}\Pi_{\mu}+m_{0}]\Psi_R(t,{\bf r}),
\label{mass3}\fe where we used $\Psi(t,{\bf r})=\Psi_L(t,{\bf r})+\Psi_R(t,{\bf r})$. 

Substituting the spinor \eqref{mass3} into Eq. \eqref{mass1}, we obtain
\ie (\gamma^{\mu}\Pi_{\mu}-m_{0})(\gamma^{\mu}\Pi_{\mu}+m_{0})\Psi_R(t,{\bf r})=0.
\label{mass4}\fe

We now consider the coupling of the spinor with the Dirac oscillator potential through the non-minimal substitution ${\bf p}\rightarrow{\bf p}-im_{0}\omega\beta{\bf r}$. In the polar coordinates system $(t,\rho,\theta)$ the metric tensor is given by $g^{\mu\nu}$=diag$(1,-1,-\rho^2)$, being $\rho=\sqrt{x^2+y^2}$ the radial coordinate and $0\le\theta\le2\pi$ the azimutal coordinate, the Eq. \eqref{mass4} becomes \cite{Villalba}
\ie \Pi^{-}\Pi^{+}\Psi_R(t,\rho,\theta)=0,
\label{mass5}\fe
where the operators $\Pi^{\mp}$ are defined in the form
\ie \Pi^{\mp}=\left[\beta\Pi_{0}+\gamma^{\rho}\left(i\frac{\partial}{\partial\rho}+im_{0}\omega\beta\rho\right)+\gamma^{\theta}\left(\frac{i}{\rho}\frac{\partial}{\partial\theta}+qA_{\theta}\right)\mp m_{0}\right],
\label{operators2}\fe with \ie \gamma^{0}=\beta,\ \gamma^{\rho}=\boldsymbol{\gamma}\cdot\hat{e}_{\rho}=\gamma^{1}\cos\theta+\gamma^{2}\sin\theta, \ \gamma^{\theta}=\boldsymbol{\gamma}\cdot\hat{e}_{\theta}=-\gamma^{1}\sin\theta+\gamma^{2}\cos\theta.
\label{mass6}\fe

Here, we are explicitly assuming that the radial component of the vector potential is null ($A_{\rho}=0$). Also, since we are work with a two-component spinor is convenient to define the Dirac matrices $\boldsymbol{\gamma}=(\gamma^1,\gamma^2)=(-\gamma_1,-\gamma_2)$ and $\beta$ in terms of the $2\times 2$ Pauli matrices, i.e., $\gamma_1=\sigma_3\sigma_1$, $\gamma_2=\sigma_3\sigma_2$ and $\beta=\sigma_3$ \citep{Greiner}. Besides that, through a similarity transformation given by unitary operator $U(\theta)=e^{-\frac{i\theta\sigma_3}{2}}$, we can reduce the rotated Dirac matrices $\gamma^{\rho}$ and $\gamma^{\theta}$ to fixed matrices $\gamma^{1}$ and $\gamma^{2}$ as follows \cite{Villalba}:
\ie U^{-1}(\theta)\gamma^{\rho}U(\theta)=\gamma^{1}, \  U^{-1}(\theta)\gamma^{\theta}U(\theta)=\gamma^{2}.
\label{mass7}\fe

Taking into account the above relations, we can rewrite the Eq. \eqref{mass5} in the form
\ie \bar{\Pi}^{-}\bar{\Pi}^{+}\psi_R(t,\rho,\theta)=0,
\label{mass8}\fe
where 
\ie \bar{\Pi}^{\mp}=\left[\sigma_3\Pi_{0}+\sigma_2\left(\frac{\partial}{\partial\rho}+m_{0}\omega\rho\sigma_3\right)+i\sigma_1\left(\frac{i}{\rho}\frac{\partial}{\partial\theta}+qA_{\theta}+\frac{\sigma_3}{2\rho}\right)\mp m_{0}\right],
\label{operators3}\fe
and
\ie \psi_R(t,\rho,\theta)=U^{-1}(\theta)\Psi_R(t,\rho,\theta).
\label{spinor1}\fe

Let us now consider a homogeneous external magnetic of intensity $B$ applied in the $z$-direction perpendicular to the $xy$ plane given by ${\bf B}_{I}(z)={\bf\nabla}\times{\bf A}_{I}=B(x,y)\hat{e}_{z}$, where ${\bf A}_{I}(x,y)=\frac{B}{2}(-y\hat{e}_{x}+x\hat{e}_{y})$. In polar coordinates, the scalar $A_{0}$ and vector ${\bf A}_{I}$ potentials are given as follows \cite{Villalba}
\ie A_0(\rho)=\frac{Z\vert e \vert}{\rho}, \ \, {\bf A}_{I}(\rho,\theta)=\frac{B\rho}{2}\hat{e}_{\theta}.
\label{potentials}\fe

Consequently, the operator $\Pi_{0}$ is written as
\ie \Pi_{0}(\rho)=i\frac{\partial}{\partial t}-qA_0(\rho)=i\frac{\partial}{\partial t}+\frac{Z\vert e \vert^{2}}{\rho},
\label{potential1}\fe
where $qA_{0}=U(\rho)=-\frac{Z\vert e \vert^{2}}{\rho}$ is the 2D Coulomb energy potential associated with a point atomic nucleus of nuclear charge $Z\vert e \vert$, where $Z$ is the atomic number. For the AB effect, the vector potential is given in the form \cite{Aharonov}
\ie {\bf A}_{II}(\rho,\theta)=\frac{\Phi_{AB}}{\rho}\hat{e}_{\theta}, \ (\rho>a),
\label{potential2}\fe
where $\Phi_{AB}\equiv\Phi/2\pi$, being $\Phi=\pi a^{2}\mathcal{B}$ the magnetic flux within a solenoid of radius $a$.

From the azimuthal components of vector potential in \eqref{potentials} and \eqref{potential2} together with the operator \eqref{potential1}, we transform the Eq. \eqref{mass8} into the following form
\ie \left[\frac{d^{2}}{d\rho^{2}}+\frac{1}{\rho}\frac{\partial}{\partial\rho}-\frac{1}{4\rho^2}-\frac{{\bf\Gamma}({\bf\Gamma}-1)}{\rho^{2}}-m_{o}^{2}\bar{\omega}^{2}\rho^{2}+\frac{{\bf\Delta}}{\rho}+{\bf\Sigma}\right]\psi_R(t,\rho,\theta)=0
\label{mass9}, \fe
where we define the following operators
\ie {\bf\Gamma}^{2}\equiv{L^{2}_z-Z^2\vert e\vert^4+\vert e \vert^2\Phi^{2}_{AB}+2\vert e \vert L_z\Phi_{AB}}, \ \ {\bf\Gamma}\equiv{iL_z\sigma_2\sigma_1+Z\vert e\vert^2\sigma_3\sigma_2+i\vert e \vert\Phi_{AB}\sigma_2\sigma_1}
\label{operators4}, \fe
\ie {\bf\Delta}\equiv{2iZ\vert e \vert^2\frac{\partial}{\partial t}}, \ \ {\bf\Sigma}\equiv{-\frac{\partial^2}{\partial t^2}-m^{2}_{0}+2m_{0}\bar{\omega}\vert e \vert\Phi_{AB}+m_{0}\bar{\omega}(2L_z+\sigma_3)},\ \ L_z=-i\frac{\partial}{\partial\theta}
\label{operators5}, \fe
being $\bar{\omega}$ an effective angular frequency given by $\bar{\omega}=(\omega-\frac{\omega_{c}}{2})\geqslant{0}$, where $\omega_{c}=\vert e \vert B/m_{0}$ is the cyclotron frequency of the particle with charge $q=-\vert e \vert$.

Now, we can separate the polar variables by taking the two-component Dirac spinor as \cite{Villalba}
\ie \psi_{R}(t,\rho,\theta)=\frac{e^{i(m_{l}\theta-Et)}}{\sqrt{2\pi\rho}}\left(
           \begin{array}{c}
            \phi^{+}(\rho) \\
             \phi^{-}(\rho) \\
           \end{array}
         \right),  \ (m_{l}=\pm1/2,\pm3/2,\ldots)
\label{spinor2}.\fe

From the Eqs. \eqref{operators4}, \eqref{operators5} and \eqref{spinor2}, it can be shown that the radial equation associated with the Eq. \eqref{mass9} takes the form
\ie \left[\frac{d^{2}}{d\rho^{2}}-\frac{\gamma(\gamma-s)}{\rho^{2}}-m_{o}^{2}\bar{\omega}^{2}\rho^{2}+\frac{2Z\vert e \vert^{2}E}{\rho}+E^{s}\right]\phi^{s}(\rho)=0
\label{mass10}, \fe
where
\ie \gamma\equiv\sqrt{(m_{l}^{2}+\vert e \vert\Phi_{AB})^2-Z^{2}\vert e \vert^{4}}
\label{mass11}, \fe
and
\ie E^{s}\equiv\left(E^{2}-m_{0}^{2}+2m_{0}\bar{\omega}\vert e \vert\Phi_{AB}+m_{0}\bar{\omega}(2m_{l}+s)\right)
\label{mass12}, \fe
with $\phi^{s}(\rho)$ being the real radial functions, $E$ is the total relativistic energy of the particle, $m_{l}$ is the orbital magnetic quantum number and $s=\pm 1$ is a spin parameter which characterizes both the two components of Dirac spinor, and it arises from the relation: $\sigma_{3}\psi_{R}=s\psi_{R}$. It is worth noting that for $\Phi_{AB}=Z=0$, the Eq. \eqref{mass10} is reduced to DO in the presence of a magnetic field \cite{Villalba}. Now, for $\Phi_{AB}=\bar{\omega}=0$, the Eq. \eqref{mass10} is reduced to the (2+1)-dimensional relativistic hydrogen atom \cite{Dong}.

\section{Bound-state solutions and energy spectrum \label{sec3}}

Here, we apply the basic concepts of the biconfluente Heun equation (BHE) formalism \cite{Vitoria,Medeiros,Bakke,Ronveaux} to deal with Eq. \eqref{mass10}. First, let us define a new coordinate in the form $x=\sqrt{m_{0}\bar{\omega}}\rho$. So we can rewrite the Eq. \eqref{mass10} as 
\ie \left[\frac{d^{2}}{dx^{2}}-\frac{\gamma(\gamma-s)}{x^{2}}-x^{2}+\frac{\bar{A}}{x}+\bar{E}^{s}\right]\phi^{s}(x)=0
\label{solution1}, \fe
where
\ie \bar{A}\equiv\frac{2Z\vert e\vert^{2}E}{\sqrt{m_{0}\bar{\omega}}}, \ \  \bar{E}^{s}\equiv\frac{E^{s}}{m_{0}\bar{\omega}}
\label{solution2}.\fe

The asymptotic behavior of the Eq. \eqref{solution1} allows us to write the function $\phi^{s}(x)$ in terms of an unknown function $f^{s}(x)$ as follows
\ie \phi^{s}(x)=C^{s}x^{\vert \gamma\vert+(1-s)/2}e^{-\frac{x^{2}}{2}}f^{s}(x),
\label{solution3}\fe 
where $C^{s}$ is a normalization constant and $\phi^{s}(x)$ must satisfy the following boundary conditions to be a physically acceptable solution: $\phi^{s}(x\to{\infty})=0$ and $\phi^{s}(x\to{0})=0$.

Substituting now the function \eqref{solution3} into Eq. \eqref{solution1}, we obtain
\ie \left[\frac{d^{2}}{dx^{2}}+\left(\frac{\delta^{s}}{x}-2x\right)\frac{d}{dx}+\left(\mathcal{E}^{s}+\frac{\bar{A}}{x}\right)\right]f^{s}(x)=0
\label{solution4},\fe
where
\ie \delta^{s}\equiv(2\vert \gamma\vert+1-s),\ \mathcal{E}^{s}\equiv\left(\bar{E}^{s}-2\vert \gamma \vert-(2-s)\right)
\label{solution5}.\fe

The Eq. \eqref{solution4} is known in the literature as the biconfluent Heun equation (BHE)  \cite{Vitoria,Medeiros,Bakke,Ronveaux} and the function $f^{s}(x)$ is the biconfluent Heun function (BHF):
\ie f^{s}(x)=H_B\left(\delta^{s}-1,0,\delta^{s}+1+\mathcal{E}^{s},2\bar{A},-x\right)
\label{solution6}.\fe

To proceed with our discussion about bound states, let us apply the Frobenius method \cite{Arfken}. Thus, the general solution to Eq. \eqref{solution4} can be written as a power series expansion around the origin:
\ie f^{s}(x)=\sum_{k=0}^{\infty} a^{s}_k x^k
\label{solution7}.\fe

Substituting the series (\ref{solution7}) into Eq. (\ref{solution4}), we obtain the following recurrence relation:
\ie a^{s}_{k+2}=-\frac{\bar{A}}{(k+2)(k+1+\delta^{s})}a^{s}_{k+1}+\frac{2k-\mathcal{E}^{s}}{(k+2)(k+1+\delta^{s})}a^{s}_k
\label{solution8},\fe together with the two independent coefficients 
\ie a^{s}_1=-\frac{\bar{A}}{\delta^{s}},
\label{solution9}\fe
\ie a^{s}_2=\frac{\bar{A}^{2}-\delta^{s}\mathcal{E}^{s}}{2\delta^{s}(1+\delta^{s})}.
\label{solution10}\fe

By starting with $a^{s}_0=1$ and using the relation (\ref{solution8}), we can calculate other coefficients of the power series expansion (\ref{solution7}). As examples, the coefficient $a^{s}_3$ is give by:
\ie a^{s}_3=-\frac{\bar{A}a^{s}_2-(2-\mathcal{E}^{s})a^{s}_1}{3(2+\delta^{s})}
\label{solution11}.\fe

As the relativistic bound state solutions require that the function (\ref{solution3}) must be normalizable, implies that the solutions of the system can be achieved by imposing that the power series expansion \eqref{solution7} or the biconfluent Heun series becomes a polynomial of degree $n$ \cite{Vitoria,Medeiros,Bakke}. Through the recurrence relation (\ref{solution8}), we can see that the power series expansion (\ref{solution7}) becomes a polynomial of degree $n$ by imposing two conditions in the form \cite{Vitoria,Medeiros,Bakke}
\ie \mathcal{E}^{s}=2n, \ \ a^{s}_{n+1}=0, \ \ (n=1,2,3,\ldots) 
\label{solution12}.\fe

Now, substituting the relation \eqref{solution12} in \eqref{solution5} and using the relations \eqref{mass12} and \eqref{solution2}, we obtain the following discrete energy spectrum of the DO in the presence of a homogeneous magnetic field in an ABC system
\ie E_{n,m_l,s}=\pm\sqrt{m_{0}^{2}+2m_{0}\bar{\omega}\left(n+\vert \gamma \vert+1-s-m_l-\vert e \vert\Phi_{AB}\right)},
\label{solution13}\fe
where $n=1,2,3,\ldots$ is the principal quantum number, the sign $+1$ corresponds the positive energies states for the particle and the sign $-1$ corresponds the negative energies states for the antiparticle.

We observed that in the case where the angular frequency of the DO is tuned to resonate perfectly with the half-frequency of the cyclotron, i.e., $\omega =\omega_c/2$, the spectrum \eqref{solution13} is reduced to the rest energy of the particle and free antiparticle. It is also worth emphasizing that for $\omega=0$, we must make the following change in the energy spectrum \eqref{solution13} $\bar{\omega}\to{\vert\omega_c\vert}$. Otherwise, we would have imaginary energies.

Now we analyse the condition $a^{s}_{n+1}=0$ given in Eq. \eqref{solution12}. To this aim, we consider the angular frequency $\bar{\omega}$ to be adjusted in such a way that $a^{s}_{n+1}=0$ can be satisfied. As a consequence this condition, we have an explicit dependency of the effective angular frequency $\bar{\omega}$, or, angular frequency $\omega$ of the DO with the quantum numbers of the system and of the spin parameter $\{{n, m_l,s}\}$. It is worth mentioning that the Klein–Gordon oscillator subject to Coulomb-type potentials was investigated in Refs. \cite{Vitoria}, and the values of the angular frequency also dependent on the quantum numbers of the system. Let us exemplify the discussion about the condition $a^{s}_{n+1}=0$ by considering first the ground state of the system $n=1$, where we have $a_2=0$. Then, using $a_2$ given by \eqref{solution10} and the expressions \eqref{solution5} and \eqref{solution2}, we obtain the following expression for the angular frequency of the DO for the ground state
\ie \omega_{1,m_{l},s}=\frac{\omega_c}{2}+\frac{2Z^2\vert e \vert^{4}E^{2}_{1,m_l,s}}{m_{0}(2\vert \gamma \vert+1-s)}
\label{solution14}.\fe

As can be seen in the expression \eqref{solution14}, besides the angular frequency of the DO be a real and positive quantity, it increases quadratically with the energy and linearly with $\omega_c$. We also observe that for $\omega=\omega_c/2 $, the expression \eqref{solution14} implies in $Z=0$, i.e., the case of the particle and free antiparticle. By substituting the angular frequency \eqref{solution14} in the spectrum \eqref{solution13}, the allowed energies levels for the ground state of the system is given by
\ie E_{1,m_{l},s}=\pm\frac{m_0}{\sqrt{1-4Z^2\vert e \vert^{4}\left(\frac{2+\vert \gamma \vert-s-m_l-\vert e \vert\Phi_{AB}}{2\vert \gamma \vert+1-s}\right)}}
\label{solution15}.\fe

In view of this, we see that the energy of the ground state is independent of the $\omega_c$ (or the external magnetic field). We will now get the eigenfunction of the ground state, which corresponds to the simplest case of the function $f^{s}(x)$ is a polynomial the first degree.  In this way, for $n=1$, we can write $f^{s}_{1,m_l}(x)=1+(\frac{\bar{A}}{\delta^{s}})x$. Thereby, the radial function \eqref{solution3} associated with the ground state is given in the form
\ie \phi^{s}_{1,m_l}(x)=C^{s}x^{\vert \gamma\vert+(1-s)/2}e^{-\frac{x^{2}}{2}}\left(1+\frac{\bar{A}}{\delta^{s}}x\right)
\label{solution16}.\fe

Now, let us consider the first excited state in which $n=2$. From the condition $a_{n+1}=0$ we have $a_3=0$, so, using \eqref{solution11} and the relations \eqref{solution2},  \eqref{solution5}, \eqref{solution9} and \eqref{solution10} we have the following possible values of the angular frequency of the DO for the  first excited state:
\ie \omega_{2,m_{l},s}=\frac{\omega_c}{2}+\frac{Z^2\vert e \vert^{4}E^{2}_{2,m_l,s}}{m_{0}(4\vert \gamma \vert+3-2s)}
\label{solution17}.\fe

By substituting the angular frequency \eqref{solution17} in the spectrum \eqref{solution13}, the allowed energies levels for the first excited state of the system is given by
\ie E_{2,m_{l},s}=\pm\frac{m_0}{\sqrt{1-2Z^2\vert e \vert^{4}\left(\frac{3+\vert \gamma \vert-s-m_l-\vert e \vert\Phi_{AB}}{4\vert \gamma \vert+3-2s}\right)}}
\label{solution18}.\fe

We will now get the eigenfunction of the first excited state, which corresponds to the polynomial of the second degree. In this way, for $n=2$, we can write $f^{s}_{2,m_l}(x)=1+(\frac{\bar{A}}{\delta^{s}})x+(\frac{\bar{A}^{2}-4\delta^{s}}{2\delta^{s}(1+\delta^{s})})x^{2}$. Thereby, the radial function \eqref{solution3} associated with the first excited state is given in the form
\ie \phi^{s}_{2,m_l}(x)=C^{s}x^{\vert \gamma\vert+(1-s)/2}e^{-\frac{x^{2}}{2}}\left(1+\frac{\bar{A}}{\delta^{s}}x+\frac{\bar{A}^{2}-4\delta^{s}}{2\delta^{s}(1+\delta^{s})}x^{2}\right)
\label{solution19}.\fe

Then, starting from the fact that the angular frequency of the DO for the $n$-th energy level of the system can be determined by condition \eqref{solution12}, coefficients \eqref{solution9}, \eqref{solution10} and \eqref{solution11} and recurrence relation \eqref{solution8}, the energy spectrum \eqref{solution13} is rewritten in a more general form as
\ie E_{n,m_l,s}=\pm\sqrt{m_{0}^{2}+2m_{0}\bar{\omega}_{n,m_l,s}\left(n+\vert \gamma \vert+1-s-m_l-\vert e \vert\Phi_{AB}\right)}.
\label{solution20}\fe

Finally, we can conclude that the ground state of the DO becomes defined by the quantum number $n=1$ instead of the quantum number $n=0$ as obtained in Ref. \cite{Villalba}. Besides, the angular frequency of the DO is determined by the quantum numbers $n$ and $m_l$ and the spin parameter $s$, whose meaning of this condition is that only specific values of the angular frequency $\omega$ are allowed, so that the relativistic bound state solutions can be achieved.

\section{Conclusion\label{conclusion}}

In this work, we have studied the (2+1)-dimensional Dirac oscillator in the presence of a homogeneous magnetic field in an Aharonov-Bohm-Coulomb system. We applied the projections operators  $left$-$handed$ and $right$-$handed$ into the Dirac oscillator equation to obtain a second-order differential equation. Assuming an ansatz to the solution which ensures a physically acceptable asymptotic behavior at $x\to{0}$ and $x\to{\infty}$, we obtain a biconfluent Heun equation. From this result, we found the energy spectrum for the bound states of the system. We observe that the energy spectrum depends on the quantum numbers $n=1,2,\cdots$ and $m_l=\pm{1/2},\pm{3/2},\cdots$, the cyclotron frequency $\omega_c=\vert e \vert B/m_0$ and the parameters  $Z$ and $\Phi_{AB}$ that characterize the Aharonov-Bohm-Coulomb system. We also explicitly determine the eigenfunctions, energy levels and angular frequencies of the Dirac oscillator for the ground state ($n=1$) and the first excited state ($n=2$), where we observed that these energy levels do not depend on the external magnetic field, and the angular frequencies are real and positive and increase quadratically with the energy and linearly with $\omega_c$.

\section*{Acknowledgments}

\hspace{0.5cm}The authors would like to thank the Funda\c{c}\~{a}o Cearense de Apoio ao Desenvolvimento Cient\'{\i}fico e Tecnol\'{o}gico (FUNCAP)(PNE-0112-00061.01.00/16), and the Conselho Nacional de Desenvolvimento Cient\'{\i}fico e Tecnol\'{o}gico (CNPq) for grants Nos. 305678/2015-9 (RVM), and 308638/2015-8 (CASA).

\end{document}